\documentclass[aps,preprint,amsfonts,amssymb,amsmath,%
tightenlines,nofootinbib,showpacs,showkeys]{revtex4}

\newcommand{\cC}{\mathcal{C}}
\newcommand{\cP}{\mathcal{P}}
\newcommand{\cQ}{\mathcal{Q}}
\newcommand{\cT}{\mathcal{T}}
\newcommand{\fD}{\mathfrak{D}}
\newcommand{\fF}{\mathfrak{F}}
\newcommand{\fL}{\mathfrak{L}}
\newcommand{\fR}{\mathfrak{R}}
\newcommand{\fS}{\mathfrak{S}}
\newcommand{\bH}{\mathbf{H}}
\newcommand{\bK}{\mathbf{K}}
\newcommand{\bbR}{\mathbb{R}}
\newcommand{\bbC}{\mathbb{C}}
\newcommand{\rmi}{\mathrm{i}}
\newcommand{\rmd}{\mathrm{d}}

\newcommand{\limaint}{\lim_{a\to\infty}\int_{-a}^{a}}
\newcommand{\Gaminf}{\Gamma_{\infty}}
\newcommand{\Ker}{\operatorname{Ker}}
\newcommand{\im}{\operatorname{Im}}
\newcommand{\diag}{\operatorname{diag}}

\begin{document}



%
%

\title{$\cP\cT$-Symmetric Quantum Theory Defined in a {K}rein Space}
\author{Toshiaki Tanaka}
\email{ttanaka@mail.tku.edu.tw}
\affiliation{Department of Physics, Tamkang University,\\
 Tamsui 25137, Taiwan, R.O.C.}


\begin{abstract}

We provide a mathematical framework for $\cP\cT$-symmetric quantum
theory, which is applicable irrespective of whether a system is defined
on $\bbR$ or a complex contour, whether $\cP\cT$ symmetry is
unbroken, and so on. The linear space in which $\cP\cT$-symmetric
quantum theory is naturally defined is a Krein space constructed
by introducing an indefinite metric into a Hilbert space composed
of square integrable complex functions in a complex contour. We show
that in this Krein space every $\cP\cT$-symmetric operator is
$\cP$-Hermitian if and only if it has transposition symmetry as well,
from which the characteristic properties of the $\cP\cT$-symmetric
Hamiltonians found in the literature follow. Some possible ways to
construct physical theories are discussed within the restriction to
the class $\bK(\bH)$.

\end{abstract}






\maketitle


Since Bender and Boettcher claimed that the reality of the spectrum
of the Hamiltonian $H=p^{2}+x^{2}+\rmi x^{3}$ is due to the underlying
$\cP\cT$ symmetry \cite{BB98a}, there have appeared in the literature
numerous investigations into various aspects of non-Hermitian
Hamiltonians defined on, in general, a complex contour.
By a simple argument, the eigenvalues of any
$\cP\cT$-symmetric linear operator are shown to be real unless the
corresponding eigenvectors break $\cP\cT$ symmetry \cite{BBM99}.
The examinations in the first four years (1998--2001) were mostly
devoted to check this spectral property. For an extensive
bibliography, see e.g. the references cited in Ref.~\cite{Zn02}.
Then Dorey \textit{et al.} achieved the celebrated rigorous
proof of a sufficient condition for the spectral reality of a
multi-parameter family of a $\cP\cT$-symmetric Hamiltonian
\cite{DDT01a}.

Around the same period, the researchers in the field were gradually
interested in the other important problems such as inner products,
Hilbert spaces, completeness of the eigenvectors, and so on. These
problems were already noticed and accessed in a couple of the earlier
works \cite{FGRZ98,BCMS99,BBS00}, where a bilinear non-Hermitian form
was introduced as a metric. Then the different groups arrived
at the same sesquilinear Hermitian but indefinite form defined
in the real line $\bbR$ \cite{Ja02,BQZ01}, which was the origin of
what has been sometimes called the \emph{$\cP\cT$ inner product}.
In particular, it was discussed in Ref.~\cite{Ja02} that the state
vector space with this indefinite metric is a Krein space and that
the usual quantum mechanical description would be obtained for
$\cP\cT$-symmetric Hamiltonians as far as we `ignore' the neutral
eigenvectors. Some mathematical results of Krein space were also
applied to a simple $\cP\cT$-symmetric model defined in a finite
real interval $[-1,1]\subset\bbR$ in Ref.~\cite{LT04}.
An apparent drawback of their metric is that it was
defined only for wave functions of $L^{2}(\bbR)$. Regarding the
indefiniteness, Bender \textit{et al.} in 2002 proposed a new
operator $\cC$, called \emph{charge conjugation}, to construct
a positive definite inner product for an unbroken $\cP\cT$-symmetric
model \cite{BBJ02}, expecting that a physically acceptable quantum
theory would be obtainable with it. However, the $\cC$
operator depends on the Hamiltonian under consideration, and
the explicit construction of the $\cC$ operator has been one of
the current main issues, see e.g., \cite{BMW03,BBJ04a,BJ04}.

On the other hand, Mostafazadeh employed the notion of
\emph{pseudo-Hermiticity} and tried to formulate $\cP\cT$-symmetric
theory within its framework \cite{Mo02a}. For the development,
see Ref.~\cite{Mo05c} and the references cited therein.
However, the formulation has involved a defect from the beginning,
since the (reference) Hilbert space is basically defined in $\bbR$
and thus the formulation cannot be directly applied to the case where
a $\cP\cT$-symmetric operator is naturally defined on a complex
contour. This problem was recently addressed for a couple of models
in Ref.~\cite{Mo05a}. Besides this problem, we should call
an attention to the more serious fact that until now little has been
known about what kinds of (non-normal) pseudo-Hermitian operators in
infinite-dimensional spaces certainly guarantee one of the key
assumptions in the series of the papers, namely, the existence of
a complete biorthonormal eigenbasis of the operators,
as was questioned in Refs.~\cite{KS04,GT06}.
In this respect, we have presented a set of necessary conditions
for the existence of biorthonormal eigenbasis of non-Hermitian
operators in our previous paper~\cite{Ta06c}. Another important
caution about pseudo-Hermiticity, namely, boundedness of metric
operators, was recalled in Ref.~\cite{KS04}; see also
Ref.~\cite{SGH92}.

In this letter, considering the present status in the field,
we would like to propose a unified mathematical framework for
$\cP\cT$-symmetric quantum theory. Here by `unified' we mean that
its applicability does not rely on whether a theory is defined
on $\bbR$ or a complex contour, on whether $\cP\cT$ symmetry is
unbroken, and so on. Furthermore, we clarify the relation between
$\cP\cT$ symmetry and pseudo-Hermiticity in our framework. We then
discuss some possibilities for constructing physical theories within
our framework based on mathematically well-established results.


To begin with, let us introduce a complex-valued smooth function
$\zeta(x)$ on the real line $\zeta :\bbR\to\bbC$ satisfying that (i)
the real part of $\zeta(x)$ is monotone increasing in $x$ and
$\Re\zeta(x)\to\pm\infty$ as $x\to\pm\infty$, (ii) the first derivative
is bounded, i.e., $(0<)|\zeta'(x)|<C(<\infty)$ for all $x\in\bbR$, and
(iii) $\zeta(-x)=-\zeta^{\ast}(x)$ where $\ast$ denotes complex
conjugate. The function $\zeta(x)$ defines a complex contour in the
complex plane and here we are interested in a family of the complex
contours $\Gamma_{a}\equiv\{\zeta(x)|x\in(-a,a),\ a>0\}$,
which has mirror symmetry with respect to the imaginary axis. This
family of complex contours would sufficiently cover all the support
needed to define $\cP\cT$-symmetric quantum mechanical systems.
In particular, we note that $\Gaminf$ with $\zeta(x)=x$ is just
the real line $\bbR$ on which standard quantum mechanical systems
are considered.

Next, we consider a complex vector space $\fF$ of a certain class
of complex functions and introduce a sesquilinear Hermitian form
$Q_{\Gamma_{a}}(\cdot,\cdot):\fF\times\fF\to\bbC$ on the space
$\fF$, with a given $\zeta(x)$, by
\begin{align}
Q_{\Gamma_{a}}(\phi,\psi)\equiv\int_{-a}^{a}\rmd x\,
 \phi^{\ast}(\zeta(x))\psi(\zeta(x)).
\label{eq:inner}
\end{align}
Apparently, it is positive definite, $Q_{\Gamma_{a}}(\phi,\phi)>0$
unless $\phi=0$, and thus defines an inner product on the space $\fF$.
With this inner product we define a class of complex functions
which satisfy $\lim_{a\to\infty}Q_{\Gamma_{a}}(\phi,\phi)<\infty$,
that is, the class of complex functions which are square integrable
(in the Lebesgue sense) in the complex contour $\Gaminf$ with respect
to the \emph{real} integral measure $\rmd x$. We note that this class
contains all the complex functions which are square integrable in
$\Gaminf$ with respect to the complex measure $\rmd z$
along $\Gaminf$ thanks to the property (ii) of the function
$\zeta(x)$. As in the case of $L^{2}(\bbR)$, we can show that this
class of complex functions also
constitutes a Hilbert space equipped with the
inner product $Q_{\Gaminf}(\cdot,\cdot)\equiv\lim_{a\to\infty}
Q_{\Gamma_{a}}(\cdot,\cdot)$, which is hereafter denoted by
$L^{2}(\Gaminf)$. A Hilbert space $L^{2}(\Gamma_{a})$ for
a finite positive $a$ can be easily defined by imposing a proper
boundary condition at $x=\pm a$.

Before entering into the main subject, we shall define another
concept for later purposes. For a linear differential operator $A$
acting on a linear function space of a variable $x$, $A=\sum_{n}
\alpha_{n}(x)\rmd^{n}/\rmd x^{n}$, the transposition $A^{t}$ of
the operator $A$ is defined by $A^{t}=\sum_{n}(-1)^{n}\rmd^{n}/
\rmd x^{n}\alpha_{n}(x)$.
An operator $L$ is said to have \emph{transposition symmetry} if
$L^{t}=L$. If $A$ acts in a Hilbert space $L^{2}(\Gaminf)$, namely,
$A:L^{2}(\Gaminf)\to L^{2}(\Gaminf)$, the following relation holds
for all $\phi(z),\psi(z)\in\fD(A)\cap\fD(A^{t})\subset L^{2}(\Gaminf)$:
\begin{align}
\limaint\rmd x\,\phi(\zeta(x))A^{t}\psi(\zeta(x))=\limaint
 \rmd x\, [A\phi(\zeta(x))]\psi(\zeta(x)).
\label{eq:trans}
\end{align}

With these preliminaries, we now introduce the linear parity operator
$\cP$ which performs spatial reflection $x\to -x$ when it acts on
a function of a real spatial variable $x$ as $\cP f(x)=f(-x)$.
We then define another sesquilinear form $Q_{\Gamma_{a}}
(\cdot,\cdot)_{\cP}:\fF\times\fF\to\bbC$ by
\begin{align}
Q_{\Gamma_{a}}(\phi,\psi)_{\cP}\equiv Q_{\Gamma_{a}}(\cP\phi,\psi).
\label{eq:Pinner}
\end{align}
We easily see that this new sesquilinear form is also Hermitian since
\begin{align}
Q_{\Gamma_{a}}(\psi,\phi)_{\cP}=&\,
 \int_{-a}^{a}\rmd x\,\psi^{\ast}(-\zeta^{\ast}(x))\phi(\zeta(x))
 \notag\\
=&\,\int_{-a}^{a}\rmd x'\,\psi^{\ast}(\zeta(x'))\cP\phi(\zeta(x'))
 =Q_{\Gamma_{a}}(\psi,\cP\phi)\notag\\
=&\,Q_{\Gamma_{a}}^{\ast}(\cP\phi,\psi)
 =Q_{\Gamma_{a}}^{\ast}(\phi,\psi)_{\cP},
\label{eq:HsPm}
\end{align}
where we use the Hermitian symmetry of the form (\ref{eq:inner}) as
well as the property (iii). However, it is evident that the form
(\ref{eq:Pinner}) is no longer positive definite in general. We call
the indefinite sesquilinear Hermitian form (\ref{eq:Pinner})
\emph{$\cP$-metric}.

We are now in a position to introduce the $\cP$-metric into
the Hilbert space $L^{2}(\Gaminf)$. For all $\phi(z),\psi(z)\in
L^{2}(\Gaminf)$ it is given by $Q_{\Gaminf}(\phi,\psi)_{\cP}\equiv
\lim_{a\to\infty}Q_{\Gamma_{a}}(\phi,\psi)_{\cP}$.
It should be noted that we cannot take the two limits of the integral
bounds, $a\to\infty$ and $-a\to -\infty$, independently in order to
maintain the Hermitian symmetry of the form given in Eq.~(\ref{eq:HsPm}).
Hence, in contrast with Hilbert space of ordinary quantum mechanics,
the integration in non-symmetrical region contradicts the very
definition of the $\cP$-metric.
From the definition of $\cP$ and relation (\ref{eq:HsPm}),
we easily see that the linear operator $\cP$ satisfies $\cP^{-1}=
\cP^{\dagger}=\cP$, where $\dagger$ denotes the adjoint with respect
to the inner product $Q_{\Gaminf}(\cdot,\cdot)$, and thus is a
\emph{canonical symmetry} in the Hilbert space $L^{2}(\Gaminf)$
\cite{AI89}. Hence, the $\cP$-metric turns to belong to the class of
$J$-metric and the Hilbert space $L^{2}(\Gaminf)$ equipped with
the $\cP$-metric $Q_{\Gaminf}(\cdot,\cdot)_{\cP}$ is a Krein space,
which is hereafter denoted by $L_{\cP}^{2}(\Gaminf)$. Similarly,
a Hilbert space $L^{2}(\Gamma_{a})$ with
$Q_{\Gamma_{a}}(\cdot,\cdot)_{\cP}$ is also a Krein space
$L_{\cP}^{2}(\Gamma_{a})$.

Let us next consider a linear operator $A$ acting in the Krein
space $L_{\cP}^{2}$, namely, $A:\fD(A)\subset L_{\cP}^{2}\to\fR(A)
\subset L_{\cP}^{2}$ with non-trivial $\fD(A)$ and $\fR(A)$.
The $\cP$-adjoint of the operator $A$ is such an operator $A^{c}$
that satisfies for all $\phi\in\fD(A)$
\begin{align}
Q_{\Gaminf}(\phi,A^{c}\psi)_{\cP}=Q_{\Gaminf}(A\phi,\psi)_{\cP}
 \qquad\psi\in \fD(A^{c}),
\label{eq:Padj}
\end{align}
where the domain $\fD(A^{c})$ of $A^{c}$ is determined by
the existence of $A^{c}\psi\in L_{\cP}^{2}$. It is related to
the adjoint operator $A^{\dagger}$ in the corresponding Hilbert
space $L^{2}$ by $A^{c}=\cP A^{\dagger}\cP$ with
$\fD(A^{c})=\fD(A^{\dagger})$. A linear operator $A$ is called
\emph{$\cP$-Hermitian} if $A^{c}=A$ in $\fD(A)\subset L_{\cP}^{2}$,
and is called \emph{$\cP$-self-adjoint} if
$\overline{\fD(A)}=L_{\cP}^{2}$ and $A^{c}=A$ \cite{AI89}.
Here we note that the concept of \emph{$\eta$-pseudo-Hermiticity}
introduced in Ref.~\cite{Mo02a} is essentially equivalent
to what the mathematicians have long called \emph{$G$-Hermiticity}
(with $G=\eta$) among the numerous related concepts in the field.
Therefore, in this letter we exclusively employ the latter
mathematicians' terminology to avoid confusion. Unless specifically
stated, we follow the terminology after the book \cite{AI89}.

We now consider so-called $\cP\cT$-symmetric operators in
the Krein space $L_{\cP}^{2}$. The action of the anti-linear
time-reversal operator $\cT$ on a function of a real spatial
variable $x$ is defined by $\cT f(x)=f^{\ast}(x)$,
and thus $\cT^{2}=1$ and $\cP\cT=\cT\cP$ follow. Then an operator
$A$ acting on a linear function space $\fF$ is said to
be \emph{$\cP\cT$-symmetric} if it commutes with $\cP\cT$,
$[\cP\cT, A]=\cP\cT A-A\cP\cT=0$.

To investigate the property of $\cP\cT$-symmetric operators in the
Krein space $L_{\cP}^{2}$, we first note that the $\cP$-metric
can be expressed as
\begin{align}
Q_{\Gamma_{a}}(\phi,\psi)_{\cP}=\int_{-a}^{a}\rmd x\,[\cP\cT
 \phi(\zeta(x))]\psi(\zeta(x)).
\label{eq:Pmetric2}
\end{align}
It is similar to but is slightly different from the (indefinite)
$\cP\cT$ inner product in Ref.~\cite{BBJ02}, and reduces to the one
in Refs.~\cite{Ja02,BQZ01} if $\zeta(x)=x$ with $a\to\infty$.

Let $A$ be a $\cP\cT$-symmetric operator. By the definition
(\ref{eq:Padj}), $\cP\cT$ symmetry, and Eqs.~(\ref{eq:trans}) and
(\ref{eq:Pmetric2}), the $\cP$-adjoint of $A$ reads
\begin{align}
Q_{\Gaminf}(\phi,A^{c}\psi)_{\cP}=&\,\limaint\rmd x\,[\cP\cT A
 \phi(\zeta(x))]\psi(\zeta(x))\notag\\
=&\,\limaint\rmd x\,[\cP\cT\phi(\zeta(x))]A^{t}\psi(\zeta(x))\notag\\
=&\,Q_{\Gaminf}(\phi,A^{t}\psi)_{\cP},
\label{eq:PTadj}
\end{align}
that is, $A^{c}=A^{t}$ in $\fD(A^{c})$ for an arbitrary
$\cP\cT$-symmetric operator $A$. Hence, a $\cP\cT$-symmetric
operator is $\cP$-Hermitian in $L_{\cP}^{2}$ if and only if it
has transposition symmetry as well. In particular, since every
Schr\"odinger operator $H=-\rmd^{2}/\rmd x^{2}+V(x)$ has
transposition symmetry, $\cP\cT$-symmetric Schr\"odinger operators
are always $\cP$-Hermitian in $L_{\cP}^{2}$. The latter fact
naturally explains the characteristic properties of
the $\cP\cT$-symmetric quantum systems found in the literature;
indeed they are completely consistent with the well-established
mathematical consequences of $J$-Hermitian (more precisely,
$J$-self-adjoint) operators in a Krein space \cite{AI89} with
$J=\cP$. Therefore, we can naturally consider any $\cP\cT$-symmetric
quantum system in the Krein space $L_{\cP}^{2}$, regardless of
whether the support $\Gaminf$ is $\bbR$ or not, and of whether
$\cP\cT$ symmetry is spontaneously broken or not. It should be
noted, however, that the relation between $\cP\cT$ symmetry and
$J$-Hermiticity (more generally $G$-Hermiticity) varies according
to in what kind of Hilbert space we consider operators. This is
due to the different characters of the two concepts; any kind of
Hermiticity is defined in terms of a given inner product while
$\cP\cT$ symmetry is not \cite{GT06}.\\


Before closing this letter, we shall discuss some possible ways to
construct physical quantum theories defined in the Krein space
$L_{\cP}^{2}$. First of all, it would be to some extent restrictive
to consider only operators with transposition symmetry although we
are mostly interested in Schr\"odinger operators. For operators
without transposition symmetry, $\cP\cT$ symmetry does not
guarantee $\cP$-Hermiticity. Hence, the requirement of $\cP\cT$
symmetry alone would be less restrictive as an alternative to
the postulate of self-adjointness in ordinary quantum mechanics.
Furthermore, there are several reasons that even the stronger
condition of $\cP$-self-adjointness would be unsatisfactory.
In ordinary quantum mechanics, it is crucial that an arbitrary
physical state can be expressed as a linear combination of
eigenvectors of the Hamiltonian or physical observables under
consideration. In this respect, it is important to recall the fact
that this absolutely relies on the consequences of
the self-adjointness, namely, the completeness of eigenvectors and
the existence of an orthonormal basis composed of them. Unfortunately,
however, $J$-self-adjoint operators generally guarantee neither of
them; even the system of the root vectors of a $J$-self-adjoint
operator does not generally span a dense set of the whole Krein space,
and more strikingly, completeness of the system of the eigenvectors
does not guarantee the existence of a basis composed of such vectors
\cite{AI89}. From this point of view, the so-called class $\bK(\bH)$
\cite{Az76} would be one of the most promising constraints in
defining a physical theory.

A $\cP$-self-adjoint operator $A$ of the class $\bK(\bH)$ is, roughly
speaking, such an operator for which the Krein space $L_{\cP}^{2}$
admits a $\cP$-orthogonal decomposition into invariant subspaces of
$A$ as
\begin{align}
L_{\cP}^{2}=\mathop{[\dotplus]}_{i=1}^{\kappa}\left[
 \fS_{\lambda_{i}}(A)\dotplus\fS_{\lambda_{i}^{\ast}}(A)\right]
 [\dotplus]L_{\cP}^{2\prime},
\end{align}
where $[\dotplus]$ denotes $\cP$-orthogonal direct sum,
$\kappa$ is a \emph{finite} number, $\lambda_{i}\not\in\bbR$
are normal non-real eigenvalues of $A$, and $\fS_{\lambda}(A)$ is
a subspace spanned by the root vectors corresponding to each
eigenvalue $\lambda$:
\begin{align}
\fS_{\lambda}(A)=\bigcup_{n=0}^{\infty}\Ker\bigl((A-\lambda I)^{n}
 \bigr).
\label{eq:decom}
\end{align}
Relative to the above decomposition of the space, the operator $A$
has a block diagonal form $A=\diag(A_{1},\dots,A_{\kappa},A')$,
where $A_{i}=A|_{\fS_{\lambda_{i}}\dotplus\fS_{\lambda_{i}^{\ast}}}$
and $A'=A|_{L_{\cP}^{2\prime}}$. The spectrum of the operator $A'$
is real, $\sigma(A')\subset\bbR$, and there is at most a \emph{finite}
number $k$ of real eigenvalues $\mu_{i}$ for which the eigenspaces
$\Ker(A^{(\prime)}-\mu_{i}I)$ are degenerate. We note that all
the subspaces $\fS_{\lambda}(A)$ corresponding to the non-real
eigenvalues are \emph{neutral}, that is, all the elements are
$\cP$-orthogonal to themselves. It is evident that when $\kappa=0$,
the operator $A$ has no non-real eigenvalues. However, it
does not immediately mean that $\cP\cT$ symmetry of the system is
completely unbroken since eigenvectors belonging to real eigenvalues
can break $\cP\cT$ symmetry. In this stronger sense, the class
$\bK(\bH)$ cannot characterize unbroken $\cP\cT$ symmetry perfectly,
but it can certainly exclude a pathological case where an infinite
number of neutral eigenvectors emerges. Now, a problem is how to deal
or interpret the remaining finite number of neutral eigenvectors.

One possible way to construct a physical theory is to interpret
the neutral eigenvectors belonging to non-real eigenvalues as
physical states describing unstable decaying states (and their
`spacetime-reversal' states). After a sufficiently large time
$t\to\infty$ (or $t\to-\infty$), the probability of observing
these states would be zero. Thus in the \emph{time-independent}
description it indicates that they must have zero-norm for all
$t\in(-\infty,\infty)$, which may be consistent with their
neutrality. If such an interpretation turns to be indeed possible
(though it is completely different from the traditional treatment
such as optical potentials, complex coordinate rotations, and so on),
$\cP$-Hermitian quantum theory defined in the Krein space
$L_{\cP}^{2}$ would be able to describe, in particular, a system
where stable bound states and unstable decaying states \emph{coexist},
such as nuclear and hadron systems.

For the neutral eigenvectors corresponding to real eigenvalues,
however, it seems difficult to make a reasonable physical
interpretation. A simple way to avoid this difficulty is just
to impose the additional condition $k=0$. Another natural way
of resolution is to consider the quotient space $\Ker(A-\mu_{i}I)
/\Ker_{0}(A-\mu_{i}I)$ in each degenerate sector, where
$\Ker_{0}(A-\mu_{i}I)$ is the isotropic part of $\Ker(A-\mu_{i}I)$.
This prescription is somewhat reminiscent of the BRST quantization
of non-Abelian gauge theories; the whole state vector space of
the latter systems is also indefinite and the positive definite
physical space is given by the quotient space
$\Ker\cQ_{B}/\im\cQ_{B}$, where $\cQ_{B}$ is a nilpotent BRST
charge \cite{BRS76} and $\im\cQ_{B}$ is the BRST-exact neutral
subspace of the BRST-closed state vector space $\Ker\cQ_{B}$
\cite{KO78a} (for a review see, e.g., Ref.~\cite{KO79}).
Under the condition $k=0$ or the quotient-space prescription,
the eigenvectors of $A$ is complete in the Krein space
$L_{\cP}^{2}$ if and only if $\fS_{\lambda}(A)=\Ker(A-\lambda I)$
for all eigenvalues $\lambda$ (at least for bounded $A$)
\cite{Az80}. In this case, the system of eigenvectors can
constitute an \emph{almost} $\cP$-orthonormalized basis of
$L_{\cP}^{2}$ \cite{Az80}, that is, it is the union of a finite
subset of vectors $\{f_{i}\}_{1}^{n}$ and a $\cP$-orthonormalized
subset $\{e_{i}\}_{1}^{\infty}$ satisfying
$Q_{\Gaminf}(e_{i},e_{j})_{\cP}=\delta_{ij}$ or $-\delta_{ij}$,
these two subsets being $\cP$-orthogonal to one another such
that\footnote{Precisely speaking, when we employ the quotient-space
prescription, the Krein space in Eq.~(\ref{eq:alJo}) should be read
as $\hat{L}_{\cP}^{2}=\fL_{0}^{[\perp]}/\fL_{0}$ where $\fL_{0}=
\langle\Ker_{0}(A-\mu_{i}I)\,|\,i=1,\dots,k\rangle$ is a neutral
subspace of $L_{\cP}^{2}$.}
\begin{align}
L_{\cP}^{2}=\overline{\langle f_{1},\dots,f_{n}\rangle [\dotplus]
 \langle e_{1},e_{2},\ldots\rangle}.
\label{eq:alJo}
\end{align}
As another possibility, we would like to mention about the $\cC\cP\cT$
inner product approach \cite{BBJ02}. The linear charge-conjugation
operator $\cC$ was originally introduced to obtain a positive definite
inner product for the eigenvectors of $\cP\cT$-symmetric operators
when $\cP\cT$ symmetry is unbroken. In this sense, we do not need this
kind of operator since we have already formulated the framework with
a Hilbert space from the beginning, \emph{irrespective of whether
$\cP\cT$ symmetry is spontaneously broken or not}. Nevertheless,
there could be another positive definite metric which is more
suitable for our purpose. Suppose there is a \emph{bounded},
$\cP\cT$-symmetric linear operator $\cC$ with transposition symmetry
in the Hilbert space $L^{2}$ (thus $\cC$ is $\cP$-self-adjoint).
Then a counterpart of $\cC\cP\cT$ inner product in our formulation
would be the \emph{$\cC\cP$-metric}; indeed $\cC\cP$ is self-adjoint
in $L^{2}$ and thus defines a metric, and we have
\begin{align}
Q_{\Gamma_{a}}(\cC\cP\phi,\psi)
 =\int_{-a}^{a}\rmd x\,[\cC\cP\cT\phi(\zeta(x))]\psi(\zeta(x)).
\end{align}
Then, if $\cC$ is a $\cP$-orthogonal projection such that
$\cC:L_{\cP}^{2}\to L_{\cP}^{2\prime}$ and the $\cC\cP$-metric is
positive definite in $L_{\cP}^{2\prime}$ (a trivial example is
$\cC=\diag(0,\dots,0,\cP')$ relative to the decomposition
(\ref{eq:decom}), where $\cP'=\cP|_{L_{\cP}^{2\prime}}$), the system
would be essentially a $\cP$-Hermitian operator $A'$ defined in
$L_{\cP}^{2\prime}$ now equipped with the positive definite
$\cC\cP$-metric. In contrast with the Hilbert space $L^{2}$,
the $\cC\cP$-metric cannot be determined \emph{a priori} since
it depends on the structure of the decomposition (\ref{eq:decom})
and the subspace $L_{\cP}^{2\prime}$ for each given operator.
Hence, it almost corresponds to the $\cC\cP\cT$ inner product
in the case of unbroken $\cP\cT$ symmetry.

Finally, we should note that in several models investigated so far
in the literature, there appears an infinite number of complex
conjugate pair eigenvalues. In other words, the dimension of
the neutral invariant subspace is infinite, and they do not
belong to the class of $\bK(\bH)$. Hence, such models would not be
suitable for describing some physical systems, though their
mathematical aspects are certainly interesting.

It should be also noted that the way to set up an eigenvalue problem
for a given linear operator $A$ is not unique. In this sense, our
framework presented in this letter is just one possibility. Our
premise is just the non-triviality of $\fD(A)$ and $\fR(A)$ in
$L_{\cP}^{2}$. On the other hand, the eigenvalue problems
of $\cP\cT$-symmetric polynomial-type potentials in the literature,
such as Ref.~\cite{DDT01a}, were usually set up within the framework
of Ref.~\cite{Si75} without any metric. Hence, it is interesting to
investigate the relation among the different setups of eigenvalue
problems. For a recent study, see also Ref.~\cite{Tr05}.

A generalization of the framework to many-body systems (described
by $M$ spatial variables $x_{i}$) would be straightforward by
introducing $M$ complex-valued functions $\zeta_{i}(x_{i})$ which
satisfy similar properties of (i)--(iii) with respect to each variable
$x_{i}$ ($i=1,\dots,M$).

Various physical consequences of $\cP\cT$-symmetric theory in
our framework would be reported in detail in a subsequent publication
\cite{Ta06d}.

\begin{acknowledgments}
 This work was partially supported by the National Science Council
 of the Republic of China under the grant No. NSC-93-2112-M-032-009.
\end{acknowledgments}



\bibliography{refsels}

\begin{thebibliography}{10}
\expandafter\ifx\csname url\endcsname\relax
  \def\url#1{{\tt #1}}\fi
\expandafter\ifx\csname urlprefix\endcsname\relax\def\urlprefix{URL }\fi
\providecommand{\eprint}[2][]{\url{#2}}

\bibitem{BB98a}
C.~M. Bender and S.~Boettcher, Phys. Rev. Lett. 80 (1998) 5243.
\newblock \eprint{physics/9712001}.

\bibitem{BBM99}
C.~M. Bender, S.~Boettcher, and P.~N. Meisinger, J. Math. Phys. 40 (1999) 2201.
\newblock \eprint{quant-ph/9809072}.

\bibitem{Zn02}
M.~Znojil, J. Nonlin. Math. Phys. 9, Suppl. 2 (2002) 122.
\newblock \eprint{quant-ph/0103054}.

\bibitem{DDT01a}
P.~Dorey, C.~Dunning, and R.~Tateo, J. Phys. A: Math. Gen. 34 (2001) 5679.
\newblock \eprint{hep-th/0103051}.

\bibitem{FGRZ98}
F.~M. Fern{\'a}ndez, R.~Guardiola, J.~Ros, and M.~Znojil, J. Phys. A: Math.
  Gen. 31 (1998) 10105.

\bibitem{BCMS99}
C.~M. Bender, F.~Cooper, P.~Meisinger, and V.~M. Savage, Phys. Lett. A 259
  (1999) 224.
\newblock \eprint{quant-ph/9907008}.

\bibitem{BBS00}
C.~M. Bender, S.~Boettcher, and V.~M. Savage, J. Math. Phys. 41 (2000) 6381.
\newblock \eprint{math-ph/0005012}.

\bibitem{Ja02}
G.~S. Japaridze, J. Phys. A: Math. Gen. 35 (2002) 1709.
\newblock \eprint{quant-ph/0104077}.

\bibitem{BQZ01}
B.~Bagchi, C.~Quesne, and M.~Znojil, Mod. Phys. Lett. A 16 (2001) 2047.
\newblock \eprint{quant-ph/0108096}.

\bibitem{LT04}
H.~Langer and C.~Tretter, Czech. J. Phys. 54 (2004) 1113.

\bibitem{BBJ02}
C.~M. Bender, D.~C. Brody, and H.~F. Jones, Phys. Rev. Lett. 89 (2002) 270401.
\newblock Erratum-ibid. 92 (2004) 119902, \eprint{quant-ph/0208076}.

\bibitem{BMW03}
C.~M. Bender, P.~N. Meisinger, and Q.~Wang, J. Phys. A: Math. Gen. 36 (2003)
  1973.

\bibitem{BBJ04a}
C.~M. Bender, D.~C. Brody, and H.~F. Jones, Phys. Rev. Lett. 93 (2004) 251601.
\newblock \eprint{hep-th/0402011}.

\bibitem{BJ04}
C.~M. Bender and H.~F. Jones, Phys. Lett. A 328 (2004) 102.
\newblock \eprint{hep-th/0405113}.

\bibitem{Mo02a}
A.~Mostafazadeh, J. Math. Phys. 43 (2002) 205.
\newblock \eprint{math-ph/0107001}.

\bibitem{Mo05c}
A.~Mostafazadeh, J. Math. Phys. 46 (2005) 102108.
\newblock \eprint{quant-ph/0506094}.

\bibitem{Mo05a}
A.~Mostafazadeh, J. Phys. A: Math. Gen. 38 (2005) 3213.
\newblock \eprint{quant-ph/0410012}.

\bibitem{KS04}
R.~Kretschmer and L.~Szymanowski, Phys. Lett. A 325 (2004) 112.
\newblock \eprint{quant-ph/0305123}.

\bibitem{GT06}
A.~Gonz{\'a}lez-L{\'o}pez and T.~Tanaka, J. Phys. A: Math. Gen. 39 (2006) 3715.
\newblock \eprint{quant-ph/0602177}.

\bibitem{Ta06c}
T.~Tanaka.
\newblock On existence of a biorthonormal basis composed of eigenvectors of
  non-{H}ermitian operators.
\newblock Preprint, \eprint{quant-ph/0603075}.

\bibitem{SGH92}
F.~G. Scholtz, H.~B. Geyer, and F.~J.~W. Hahne, Ann. Phys. 213 (1992) 74.

\bibitem{AI89}
T.~Ya. Azizov and I.~S. Iokhvidov, Linear {O}perators in {S}paces with
  {I}ndefinite {M}etric (John Wiley, Chichester, 1989).

\bibitem{Az76}
T.~Ya. Azizov, Ukrain. Mat. Z. 28 (1976) 293.
\newblock (in {R}ussian).

\bibitem{BRS76}
C.~Becchi, A.~Rouet, and R.~Stora, Ann. Phys. 98 (1976) 287.

\bibitem{KO78a}
T.~Kugo and I.~Ojima, Phys. Lett. B 73 (1978) 459.

\bibitem{KO79}
T.~Kugo and I.~Ojima, Prog. Theor. Phys. Suppl. 66 (1979) 1.

\bibitem{Az80}
T.~Ya. Azizov, Sov. Math. Dokl. 22 (1980) 193.
\newblock ({E}nglish translation).

\bibitem{Si75}
Y.~Sibuya, Global {T}heory of a {S}econd {O}rder {L}inear {D}ifferential
  {E}quation with a {P}olynomial {C}oefficient (North-Holland Publishing,
  Amsterdam, 1975).

\bibitem{Tr05}
T.~D. Tai, J. Phys. A: Math. Gen. 38 (2005) 3665.
\newblock \eprint{math-ph/0502009}.

\bibitem{Ta06d}
T.~Tanaka.
\newblock General aspects of {$\mathcal{PT}$}-symmetric and
  {$\mathcal{P}$}-self-adjoint quantum theory in a {K}rein space.
\newblock In preparation.

\end{thebibliography}
\bibliographystyle{npb}



\end{document}